# Digital Twin of Distribution Power Transformer for Real-Time Monitoring of Medium Voltage from Low Voltage Measurements

Panayiotis Moutis, *Senior Member, IEEE*, Omid Alizadeh-Mousavi

*Abstract*—Real-time monitoring of distribution systems has become necessary, due to the deregulation of electricity markets and the wide deployment of distributed energy resources. To monitor voltage and current at sub-cycle detail, requires, typically, major investment undertaking and disruptions to the operation of the grid. In this work, measurements of the low voltage (LV) side of distribution transformers (T/F) are used to calculate in real time the waveforms of their medium voltage (MV) sides, based on a mathematical model of said T/F. This model is, essentially, the digital twin of the MV side of the T/F. The method calculates T/F MV waveforms of voltage and current, and active and reactive power as accurately as an instrument T/F, captures all harmonics content, is unaffected by asymmetrical loading and identifies most system faults on the MV side of the T/F. The digital twin method enables monitoring of distribution T/F that avoids MV instrumentation, does not suffer in accuracy and may be readily deployable. Field data from an actual MV-LV T/F, agree with simulation results showcasing the efficacy of the digital twin method.

## NOMENCLATURE

### A. Variables and Parameters

| | |
|---|---|
| $R$ | Resistance. |
| $L$ | Inductance. |
| $u$ | Voltage. |
| $i$ | Current. |
| $t$ | Time. |
| $n, N$ | Discretized time steps and total (from sampling). |
| $f$ | Frequency. |

### B. Scripts and indices

| | |
|---|---|
| $S, M$ | Series and Shunt element/branch. |
| $1, 2$ | Primary and Secondary transformer windings. |
| $s$ | Sampling. |
| $k$ | Scenario numbering. |
| $d, r$ | Digital twin and real waveform. |
| $A, B, C$ | Three-phase phases. |

## I. INTRODUCTION

Many reasons have led to a growing interest in monitoring Distribution Systems (DSs) closely and in real time. Firstly, deployment of distributed energy resources (DER) in modern power systems, in order to electrify transportation, heating and cooling, raises multiple technical concerns [1]. All actions to handle these concerns, both preventive and corrective, are either limiting the deployment of more DER or curtailing the energy that these may exchange with the grid [2]. Online monitoring of system behavior with high DER penetrations can tackle efficiently most of the known technical concerns and others, not yet foreseen. Indicatively, system situational awareness (congestion, voltage stability, etc.) [3], use of innovative system stability controls [4] as also traditional ones [5], assessment of quality of supply, awareness of availability and timely activation of flexibility services [6], and efficient fault identification and localization are few of the contributions of real-time system monitoring in this context [7]. Secondly, online measurements of systems hosting DER and extracting metrics of average loading of the equipment will contribute positively in assessing investment deference in grid infrastructure [8]. Thus, all available grid hosting capacity can be made of use and avoid the traditional oversizing design of power systems. In a similar sense, situations that affect equipment undesirably (e.g. inverse flows on relays) may be captured and handled accordingly. Thirdly, system operators seek to improve power quality and better serve electric equipment according to performance standards [9].

### A. Distribution System and Transformer Monitoring Review

The above points have led to extensive research and innovative applications regarding monitoring medium (MV) & low voltage (LV) DSs either at the extent of whole feeders or as small as specific components in them. The methods and technologies typically employed at transmission infrastructure are showing the way; transmission lines and substations are monitored either with Supervisory Control and Data Acquisition (SCADA) systems [10] or, more recently, Phasor Measurement Units (PMU), and operators monitor the grid at sub-second time scale [11]. At DSs, much research has focused on methods of state estimation [12, 13], PMU for this level specifically (uPMU) [14], microgrid design and control and energy management systems of various purposes and capacities [15]. In actual practice, there are different levels of deployments of data acquisition and managements systems at the distribution level of grids around the world. SCADA, distribution management systems and Advanced Metering Infrastructure [16] – most recently with smart meters – have been those which have been adopted more widely, while installations and use of uPMU have

P. Moutis & O. Alizadeh-Mousavi are with DEPsys SA, Route du Verney 20B, 1070 Puidoux, Switzerland (email: panayiotis.moutis@depsys.ch; omid.mousavi@depsys.ch). This project has received funding from the European Union's Horizon 2020 program under the Marie Sklodowska-Curie grant agreement No 797451. **Preprint under Green Open Access policy** been at earlier stages. Most of these works focus at the MV level of DSs, which raises a concern of much practical interest – the use of voltage transformers for measurements [17]. Given the extent of DS feeders, the costs, risks and network disruptions (e.g. replacement of cable headend) to deploy measurement T/F, can gravely affect the goals for monitoring DS.

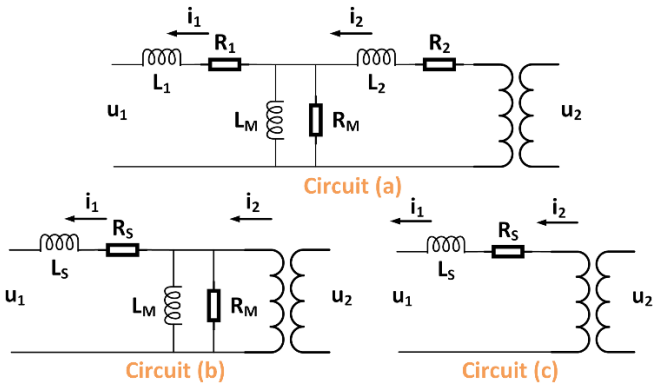

Fig. 1. Circuit models of a power T/F at various levels of detail.

From the aspect of monitoring the operation and health of specific system components, power transformers (T/F) are at the center of attention of many studies. In [18], measurements from both sides of the T/F help determine copper and iron losses under harmonics-heavy conditions. In-rush and fault currents are identified with wavelet methods in [19]. Harmonics analysis is used in [20] to capture geomagnetic disturbance on T/F. Monitoring the higher-voltage side of the T/F in [21] assists in assessing the insulation performance under lightning impulses, while frequency response analysis [22] can be an alternative to locate faults in the equipment. Maintenance and ageing concerns are also addressed more recently with greater attention [23, 24]. The value of measuring T/F performance, especially for power quality matters, reflects also in the focus on instrument T/Fs accuracy [25]. As for using uPMUs to monitor T/F, their high costs and data requirements are prohibiting factors.

### B. Innovation of the Power Transformer Digital Twin

The following proposal is put forward in this paper. The MV side of a distribution T/F may be monitored via its digital twin, instead of directly measuring its voltage and current waveforms at its terminals. The digital twin [26] of the MV side will be informed by current and voltage waveform measurements of the LV side of the distribution T/F. A digital twin is a digital replica of an actual entity, expected to behave and perform same to the actual entity. Hence, a digital twin offers insight to events that might occur to the actual entity or serve as the online model of it. The latter case is the one that this proposal is based on. The potential of digital twins in the power systems sector has only been recently acknowledged and explored [27].

To the authors' knowledge there has been no effort like the one presented in this manuscript. A digital twin of a power T/F has been proposed only as a means to assess its health status [28], but not to monitor voltage, current and power in real time. To monitor a DS MV-LV power T/F, voltage and current waveforms of the LV side are measured to calculate the respective waveforms of the MV side. Hence, the proposed work develops the digital twin of the MV side of the power T/F, based on monitoring of its LV side. The scope of this work spans DS monitoring in real time (for diagnosing faults and power quality issues) as also for long-term operating and planning assessments (feeder loading, hosting capacity, etc). The work contributes in the scoped field as follows:

- Monitoring DSs at high time granularity, as anticipated by [12-16], is made possible,
- The waveform monitoring allows to determine power quality in real time [9], as it captures all harmonics content,
- MV DS behavior under faults [7] can be captured fully to alert the system operator and logged for further analysis,
- The MV-side waveforms outputted by the digital twin of the T/F are as accurate as the measurements of an instrument T/F [17] on the MV side of the actual T/F,
- Technical personnel and system operator can assess immediately any remedial actions to system events [6],
- The cost of measurement instrumentation at the LV side of a power T/F is considerably lower than that of the MV side or of monitoring both sides, thus, making the method less costly for system operators with multiple feeders to cover,
- Installation of LV side measurement devices requires the MV network to be interrupted under fewer circumstances, thus, enabling a comparably seamless deployment,
- Works on uPMU and DS state estimation at the MV and the here proposed method may complement each other.

All the aforementioned points describe the digital twin as a method for monitoring the MV-side of a distribution T/F with limited cost, no disruption to the DS for deployment, high accuracy, and enabling fault diagnosis (previously unprecedented from monitoring the LV-side of feeders).

The innovation of this work is the real-time monitoring of the MV side of a distribution T/F without directly measuring the MV side, but by rather measuring the LV-side of said T/F.

The remainder of this work is organized as follows. Section II describes the methodology employed and the basic assumptions both for single and three phase T/F. Section III describes the testing set-up, the phenomena that are assessed and the metrics used. Section IV gathers all tests conducted and a discussion ensues. Section V uses actual LV and MV measurements from a DS T/F in Switzerland, to show that the digital twin method performs effectively and according to the statistical testing results of Section IV. Section VI concludes this work and proposes future considerations.

## II. METHODOLOGY AND REQUIRED ASSUMPTIONS

### A. Digital Twin of a Single-Phase Power Transformer

The idea of the proposed methodology is based on taking measurements of voltage and current on the low-voltage side of the T/F and, by using the mathematical description of a typical model of single-phase T/F, calculate (as an estimate/projection) what are the values of the voltage and current on the MV side of the T/F. In this sense, the MV side of the T/F is emulated *in silico*; its digital twin is implemented. As well documented, T/F can be modeled as 2-port systems; either a pi- or a t-equivalent circuit with the resistances, reactances lumped or split, according to the degree of detail sought for (see Fig. 1) [29]. Circuit model (a) is the full T/F model, circuit model (b) assumes that all series resistance and reactance is lumped on one side of the T/F as $R_s=R_1+R_2$ and $L_s=L_1+L_2$, while circuit model (c) simplifies (b) by assuming infinite impedance from $L_m$ and $R_m$. Due to its simplicity, circuit model (b) is the preferred digital twin of the MV side of the single-phase T/F for the proposed methodology. Additional detail will only improve the results that follow. As of circuit (b) and assuming that the T/F is sized and operated according to standard [30] (i.e. T/F core saturation is avoided, otherwise a piece-wise formulation [31] may complement the following set-up), the MV calculations are as:

$$\left.\begin{aligned} u_2(t) &= u'_1(t) + R_S i'_1(t) + L_S \frac{di'_1(t)}{dt} \\ i_2(t) &= \frac{u_2(t)}{R_M} + \frac{1}{L_M}\int u_2(t)\,dt + i'_1(t) \end{aligned}\right\} \quad (1)$$

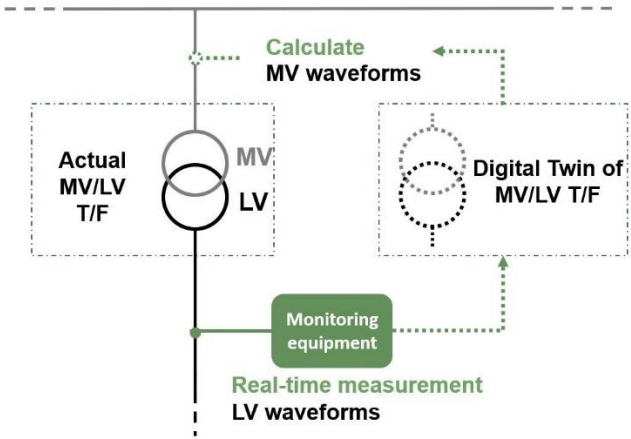

Fig. 2. Overall topology of the digital twin method for the calculation of the MV side waveforms of a DS T/F based on LV waveform measurements.

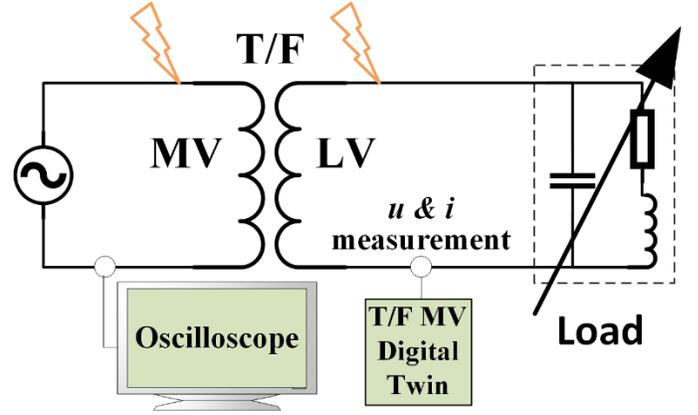

Fig. 3. Block diagram of simulation testing topology as set up in MATLAB.

Where $u$, $i$, $R$ and $L$ are voltage, current, resistance and inductance, respectively. Resistances $R_S$ and $R_M$ may also be expressed as functions of temperature or of T/F loading (i.e. inferring temperature), to allude to the effect of temperature on resistances at different loads. Voltage and current are given as time variables, since waveforms are measured. Subscripts $1$, $2$, $S$ and $M$ denote the LV and MV sides, series and shunt (leakage and magnetizing impedances) parts of the single-phase T/F, respectively. Voltage and current measurements of the LV side are referenced to the MV side (i.e. multiplied by the T/F ratio). Any tap-changing action in the T/F is considered an input to the digital twin model. Alternatively, tap-changing can be monitored electrically by the digital twin and, thus, adjust values in (1). Let it be stressed that (1) may also calculate any harmonics content either in the voltage or the current of the MV-side of the T/F, provided it is present in the LV-side measurements. The only concern with regards to this calculation stems from any filtering effects probably caused by the T/F impedance. These concerns will be thoroughly assessed in Section IV.

From (1) monitoring of the voltage and current waveforms of the LV side can yield voltage and current waveforms of the MV side of the T/F. In detail, an AC voltage and current measurement device is connected to the LV side of the single-phase T/F and measures the respective values. The voltage and current sensors provide the measurements in analog form and the measurement device converts the analog signals to digital at a sampling rate of $f_s$ (sampling interval $1/f_s$ s). With the use of the last two measurement samples, i.e. $n$ and $n$-$1$, of voltage and current at every sampling interval $n$, and via formulation (1) the voltage and current of the MV side are retrieved. Measurements $n$ and $n$-$1$ are required at every interval $n$, because of the effect of the derivative and the integral in (1). The discretized formulation of (1) is given, as the formal definition of the digital twin of the MV side of the T/F as by LV side measurements:

$$\left.\begin{aligned} u_2[n] &= u'_1[n] + R_S i'_1[n] + L_S (i'_1[n] - i'_1[n-1])f_s \\ i_2[n] &= \frac{u_2[n]}{R_M} + \frac{u_2[n] - u_2[n-1]}{L_M \cdot f_s} + i'_1[n] \end{aligned}\right\} \quad (2)$$

With regards to harmonics content, formula set (2) is different to the continuous set of (1), as the LV-side measurements must be properly sampled at high rates, so that no harmonic content is missed. This will be explored with assessment of multiple sampling rates in Section IV and consequent analysis.

### B. Digital Twin of a Three-Phase Power Transformer

For the digital twin of a three-phase T/F, the approach builds on that of the single-phase as follows: the digital twins of three single-phase T/F, each taking separate single-phase voltage and current measurements from the LV side of a three-phase T/F, are appropriately integrated to emulate the three-phase voltage and current of the MV side of the T/F. Practically, voltage and current measurements of each phase on the LV side are used to calculate the corresponding values of one of the phases on the MV side through (2). The overall digital twin topology is shown in Fig. 2. Following, the calculated values are elaborated according to the vector group of the T/F; i.e. the connection of the three phase windings. For the most commonly T/F vector groups in DSs, the phase voltages and the line currents of the MV side for one of the phases are as follows:

**Yy0**: $u_A = u_{2A}$ and $i_A = i_{2A}$ (3)
**Dy1**: $u_{AB} \cdot \sqrt{3} = u_{2A} - u_{2C}$ and $i_A = i_{2A} - i_{2C}$ (4)
**Dy11**: $u_{AB} \cdot \sqrt{3} = u_{2A} - u_{2B}$ and $i_A = i_{2A} - i_{2B}$ (5)

Where subscripts $A$, $B$, $C$ denote the three phases of the MV side of the digital twin of the three-phase T/F, and subscript $2X$ (where $X = A, B, C$) denotes the calculated values of the single phase MV side digital twin from LV measurements via (2). Grounding either T/F side does not alter formulations (3-5). The digital twin of a T/F connected as Dy will not be able to calculate MV-side current harmonics of orders multiples of the third, since such T/F topologies eliminate said harmonics [32].

## III. TESTING FRAMEWORK AND ASSESSMENT METRICS

Retrieving the MV side waveforms from monitoring the ones of the LV side of a triphase T/F has to be tested for various levels of loads connected to the T/F, changing load (increasing/decreasing), presence of harmonics in the system, non-balanced three-phase loads, non-balanced voltage supply and system faults. All these situations will affect the LV side waveforms differently and, by that, the performance of the methodology, too. A last parameter that needs to be considered is that of the parameters of the measurement device used for the application (i.e. sampling rate $f_s$). As from the above, sets of testing scenarios are defined in Subsection A.

All simulation tests are conducted in MATLAB. For most of the tests, a three-phase T/F is tested, while a single-phase T/F is used for the study of some specific concerns. The test topology (see Fig. 3) comprises a voltage source, feeding a load impedance via a MV-LV T/F – all components are modeled precisely with ratings given in the Appendix. The described topology is not unique and represents, practically, any typical DS T/F connection. This means that any of the typical IEEE or other test systems could be implemented around the described test topology of the DS T/F and the digital twin. The inclusion of any DER in the test topology would also not affect the set-up, as the direction of flow of currents does not affect the digital twin calculations. This T/F will be monitored through the proposed methodology. This T/F is the full model as found in MATLAB libraries and has not been simplified, i.e. it represents the actual T/F of the simulation tests with synthetic data. The methodology, as described in Section II, will sample the waveforms of voltage and current on the LV side of the T/F. A GridEye measurement device (class-A power quality monitoring certified) is assumed to sample the waveforms [33]. GridEye is assumed to have a random error in the interval of its accuracy (see Appendix). It will then calculate the respective voltage and current waveforms and active and reactive power values of the MV side – i.e. its digital twin behavior. The digital twin behavior will be compared for accuracy to the respective waveforms of the MV side of the actual T/F model (oscilloscope assumed connected). The metrics of accuracy used are defined in Subsection D.

### A. Statistical Testing of Digital Twin Methodology

The first set of tests is that of normal operation varied according to the following conditions:
i.  Constant load, increasing load and decreasing load; the load increase/decrease (step change) will occur at the 0.02s mark of a 0.04s simulation interval.
ii. Presence or absence of harmonics in the system at the maximum allowed level described by standards [34].
iii. Sampling rates of the measurement device at $f_s$ = [5kHz, 10kHz, 30kHz, 52kHz].

For every option in (i-iii), a different scenario is described, thus, a total of 24 scenarios is tested. The above described conditions and the values assessed are non-linear. Hence, statistical analysis will be employed [35]. This means that multiple similar tests/simulations will be conducted for every scenario. From these tests, averages, minima and maxima of the differences between actual and calculated waveforms will be calculated. For every scenario about 17000 simulations with random initial and final conditions are conducted, to ensure statistical validity of the assessment metrics (see Subsection III.D) in the 99% confidence interval at ±1% error. The statistical analysis, according to [35] ensures confidence in the performance metrics collected.

### B. Testing of Digital Twin Methodology under System Faults, Unbalanced Loading and Tap-Changing Operations

These tests will show if the accuracy of the digital twin is affected by system faults, asymmetrical behavior either at the MV or LV side of the T/F, and tap-changing action of the T/F. The variations that need to be assessed are as follows:
iv. All types of system faults (i.e. line-to-ground (LG), LL and LLG – occurring at 0.2s of 0.4s simulation), asymmetrical voltage magnitudes fed to the T/F, asymmetrical load connected to the LV side of the T/F and tap-changing action.
v. Presence/absence of a Peterson coil at the nearest substation (electrically); the said coil is commonly used to reduce ground fault currents and sustain three-phase LV operation.
vi. Grounding or not of the Y connected side of the T/F (Dy1 or Dy11 vector groups might be assessed interchangeably, grounding of both sides of Yy needs to be considered).
vii. Location of the fault at the MV or LV side of the T/F.

This set includes 72 scenarios. For each scenario a case of load about equal to the rated of the T/F is assumed to be served at the time. The waveforms of the digital twin and of the actual MV side of the T/F will be compared on similarity of behavior without the use of metrics. The reason for not using statistical metrics and as it will be noted in Section IV, is because the digital twin of the MV side is either correctly calculating the actual waveforms of the T/F or completely miscalculates one or two phases by magnitude or phase shifting. In this sense, the statistical metrics as extracted from scenarios (i-iii) will either be valid for a correct calculation by the method in scenarios (iv-vii) or the method fails substantially. All testing for the second set will be conducted at the presence of harmonics (see details in (ii) above) as this represents the worst-case scenario for the monitoring methodology here proposed.

### C. Filtering Effect of the Digital Twin on Harmonics

The last set of tests aims to assess the effect of the T/F model used (see (1),(2) and Fig. 1) to extract the digital twin of the MV side of the T/F, with regards to whether it captures all voltage and current components. The tests will be conducted on MATLAB, but for a single-phase T/F, due to simplicity. Both the actual T/F and its models, as circuits, are, essentially, low-pass filters, which damp the magnitude of higher-order harmonics. That been said, the model of the T/F used for the digital twin of its MV side might be filtering higher harmonics differently than the actual T/F, hence overestimating or underestimating the harmonic distortion at the MV level of the grid. To address this concern, firstly, the Fourier transformation of voltage and current waveforms of the digital twin and the actual MV side of the T/F will be compared for simulations of changing load at the presence of harmonics (see details in (b) above). Secondly, the Bode diagram of the model used for the digital twin and that of the circuit of the actual T/F will be compared for high and low loading of the T/F.

TABLE I
STATISTICS OF $\widetilde{\Delta x}_t$ FOR NORMAL OPERATION WITHOUT HARMONICS, GIVEN IN % OF RMS VALUE OF CORRESPONDING VARIABLE, FOR CONSTANT (0), INCREASING (+) AND DECREASING (-) LOAD IMPEDANCE

| $f_s$ (kHz) | $\Delta L$ | 0 Avg | 0 Max | 0 Min | + Avg | + Max | + Min | - Avg | - Max | - Min |
|---|---|---|---|---|---|---|---|---|---|---|
| 5 | V | 3.7 | 7.4 | 3.4 | 3.8 | 5.4 | 3.5 | 3.8 | 5.4 | 3.5 |
| | I | 2.7 | 5.6 | 1.8 | 2.6 | 4.4 | 2.0 | 2.6 | 4.4 | 2.0 |
| | P | 4.9 | 36.4 | 0.01 | 2.4 | 7.6 | 0.1 | 2.4 | 8.5 | 0.1 |
| | Q | 8.6 | >100 | 1.5 | 3.1 | 29.2 | 1.5 | 3.1 | 60.7 | 1.6 |
| | $f_V$ | 0.0 | 0.0 | 0.0 | 0.004 | 0.05 | 0.0 | 0.004 | 0.05 | 0.0 |
| | $f_I$ | 0.0 | 0.0 | 0.0 | 0.02 | 0.08 | 0.0 | 0.02 | 0.08 | 0.0 |
| 10 | V | 1.9 | 4.7 | 1.6 | 2.1 | 3.8 | 1.7 | 2.1 | 3.5 | 1.7 |
| | I | 2.5 | 4.9 | 0.9 | 1.6 | 3.3 | 0.9 | 1.6 | 3.4 | 0.9 |
| | P | 4.8 | 38.7 | 0.01 | 2.4 | 7.6 | 0.1 | 2.3 | 9.8 | 0.1 |
| | Q | 7.0 | >100 | 1.5 | 3.0 | 29.2 | 1.5 | 3.5 | 50.4 | 1.5 |
| | $f_V$ | 0.0 | 0.0 | 0.0 | 0.002 | 0.03 | 0.0 | 0.002 | 0.03 | 0.0 |
| | $f_I$ | 0.0 | 0.0 | 0.0 | 0.02 | 0.06 | 0.0 | 0.02 | 0.2 | 0.0 |
| 30 | V | 0.9 | 4.9 | 0.6 | 1.1 | 2.7 | 0.8 | 1.1 | 2.7 | 0.8 |
| | I | 2.9 | 5.5 | 0.3 | 1.7 | 3.0 | 0.4 | 1.7 | 3.2 | 0.5 |
| | P | 4.9 | 36.4 | 0.01 | 2.4 | 8.5 | 0.1 | 2.3 | 7.6 | 0.1 |
| | Q | 8.5 | >100 | 1.5 | 3.0 | 60.3 | 1.5 | 3.5 | 57.2 | 1.6 |
| | $f_V$ | 0.003 | 0.005 | 0.0 | 0.003 | 0.02 | 0.0 | 0.003 | 0.02 | 0.0 |
| | $f_I$ | 0.002 | 0.005 | 0.0 | 0.01 | 0.04 | 0.0 | 0.01 | 0.1 | 0.0 |
| 52 | V | 0.7 | 3.5 | 0.4 | 0.9 | 2.6 | 0.6 | 0.9 | 2.4 | 0.6 |
| | I | 3.1 | 5.6 | 0.3 | 1.8 | 3.4 | 0.4 | 1.9 | 3.2 | 0.5 |
| | P | 4.8 | 38.7 | 0.01 | 2.4 | 7.6 | 0.1 | 2.3 | 9.7 | 0.1 |
| | Q | 7.0 | >100 | 1.5 | 3.0 | 29.2 | 1.5 | 3.5 | 50.3 | 1.6 |
| | $f_V$ | 0.003 | 0.008 | 0.0 | 0.003 | 0.02 | 0.0 | 0.004 | 0.02 | 0.0 |
| | $f_I$ | 0.002 | 0.008 | 0.0 | 0.01 | 0.04 | 0.0 | 0.01 | 0.2 | 0.0 |

TABLE II
STATISTICS OF $\Delta x_{t,M}$ FOR NORMAL OPERATION WITHOUT HARMONICS, GIVEN IN % OF RMS VALUE OF CORRESPONDING VARIABLE, FOR CONSTANT (0), INCREASING (+) AND DECREASING (-) LOAD IMPEDANCE

| $f_s$ (kHz) | $\Delta L$ | 0 Avg | 0 Max | 0 Min | + Avg | + Max | + Min | - Avg | - Max | - Min |
|---|---|---|---|---|---|---|---|---|---|---|
| 5 | V | 8.8 | 14.4 | 8.3 | 9.9 | 11.5 | 8.4 | 9.1 | 11.5 | 8.4 |
| | I | 6.2 | 10.6 | 4.2 | 8.2 | 11.7 | 6.2 | 10.3 | 35.3 | 6.0 |
| | P | 4.9 | 36.7 | 0.1 | 2.6 | 8.9 | 0.2 | 2.6 | 8.9 | 0.2 |
| | Q | 8.6 | >100 | 1.5 | 4.2 | 77.3 | 1.9 | 4.2 | 77.3 | 1.9 |
| 10 | V | 4.4 | 8.5 | 3.9 | 4.8 | 7.4 | 4.1 | 4.8 | 7.0 | 4.1 |
| | I | 4.9 | 8.3 | 2.0 | 3.5 | 7.1 | 2.3 | 3.6 | 7.4 | 2.3 |
| | P | 4.8 | 39.0 | 0.1 | 2.5 | 7.9 | 0.1 | 2.7 | 10.6 | 0.2 |
| | Q | 7.0 | >100 | 1.5 | 4.0 | 40.4 | 1.9 | 4.0 | 57.8 | 1.8 |
| 30 | V | 1.9 | 7.6 | 1.4 | 2.3 | 4.7 | 1.6 | 2.3 | 4.9 | 1.7 |
| | I | 4.7 | 8.4 | 0.8 | 3.0 | 4.8 | 1.0 | 3.0 | 5.0 | 0.9 |
| | P | 5.0 | 36.7 | 0.1 | 2.6 | 8.9 | 0.1 | 2.6 | 8.3 | 0.1 |
| | Q | 8.5 | >100 | 1.5 | 3.9 | 76.5 | 1.8 | 3.9 | 71.2 | 1.8 |
| 52 | V | 1.3 | 5.4 | 0.9 | 2.0 | 4.8 | 1.2 | 1.8 | 4.2 | 1.2 |
| | I | 4.7 | 8.3 | 0.6 | 2.9 | 5.1 | 0.8 | 3.0 | 4.9 | 0.8 |
| | P | 4.8 | 39.0 | 0.1 | 2.5 | 7.9 | 0.1 | 2.6 | 10.4 | 0.2 |
| | Q | 7.0 | >100 | 1.5 | 3.9 | 40.2 | 1.8 | 3.9 | 57.7 | 1.8 |

## D. Assessment Metrics

For the first set of scenarios, regarding all options in (i-iii), the waveforms of voltage and current, values of active and reactive powers, and frequency of the digital twin of the MV side of the T/F (as calculated by the proposed methodology) will be compared to those of the actual T/F. Specifically for frequency, the zero crossing of ten consecutive periods of waveforms of current and voltage (compared separately) will be averaged over the time between the first and last crossing and the respective frequency will be calculated and compared via a similar elaboration of the actual waveform of voltage or current of the MV side. For the said comparisons, metrics of average and point errors between the actual waveforms and that of the digital twin will be used, and are defined as follows.

For one of the simulations in scenario $k$, assuming that $x_{d,k}$ is the waveform (voltage or current) of the digital twin of the MV side of the T/F and $x_{r,k}$ the respective waveform of the actual T/F on its MV side, the two will be compared point-to-point for every time sample interval $n$. Firstly, the average error of a waveform outputted by the digital twin normalized over the RMS of the actual waveform of the T/F is defined as:

$$\widetilde{\Delta x}_t = \sqrt{\frac{\sum_{n=1}^{N}(x_{d,k,n} - x_{r,k,n})^2}{x_{r,k,RMS}^2 \cdot N}} \quad (6)$$

where $N$ is the sample points taken in the simulated testing time interval of 0.4s. The maximum point error of the waveform outputted by the digital twin normalized over the RMS of the actual waveform of the T/F is calculated by:

$$\Delta x_{t,M} = max\left(\sqrt{\frac{(x_{d,k,n} - x_{r,k,n})^2}{x_{r,k,RMS}^2}}\right) \quad (7)$$

The metrics describe the error between the digital twin and the actual MV side T/F waveforms for one simulation of a scenario. For the case of a load change $\Delta L$ at the point of 0.2s of the testing interval, metrics $\widetilde{\Delta x}_t$ and $\Delta x_{t,M}$ are adjusted for samples $n$ in the time interval of $t$=0.16-0.24s, i.e. two electric cycles before and after the load change.

From all simulations of each scenario statistical measures will be extracted as here defined. The average, the maximum and minimum of metrics $\widetilde{\Delta x}_t$ and $\Delta x_{t,M}$ for all simulations of each scenario are extracted. Hence, the metrics presented following are the metrics over the entirety of simulations of each scenario tested, so as to ensure statistical validity. The analysis represents realistically the whole range of operating situations.

## IV. RESULTS AND DISCUSSION

The Section is organized in two subsections, the first one summarizing the results per sets of scenarios with common characteristics and the second one discussing the findings in light of the efficacy of the proposed methodology.

### A. Results

*1) Statistical Testing under Normal Operation:* In Tables I-II the statistics of metrics $\widetilde{\Delta x}_t$ and $\Delta x_{t,M}$ for voltage, current, active and reactive power, and system frequency are presented for the case that no harmonics are present in the grid, while Tables III-IV the same metrics at the presence of all harmonics within the limits of [34]. The statistics given are average (Avg), maximum (max) and minimum (min) for metrics $\widetilde{\Delta x}_t$ and $\Delta x_{t,M}$, accordingly. "0" denotes a scenario of constant load, "+" of increasing, and "-" of decreasing. $V$, $I$, $P$, $Q$, $f_V$ and $f_I$ are voltage, current, active and reactive power, and frequency calculated by the voltage and current waveforms, respectively. In

TABLE III
STATISTICS OF $\widetilde{\Delta x}_t$ FOR NORMAL OPERATION WITH HARMONICS, GIVEN IN % OF RMS VALUE OF CORRESPONDING VARIABLE, FOR CONSTANT (0), INCREASING (+) AND DECREASING (−) LOAD IMPEDANCE

| $f_s$ (kHz) | $\Delta L$ | 0 | | | + | | | − | | |
|---|---|---|---|---|---|---|---|---|---|---|
| | | Avg | Max | Min | Avg | Max | Min | Avg | Max | Min |
| 5 | V | 5.0 | 8.1 | 4.8 | 5.1 | 6.3 | 4.8 | 3.8 | 5.1 | 3.4 |
| | I | 2.8 | 5.7 | 1.8 | 2.5 | 4.7 | 2.0 | 9.4 | 25.3 | 2.6 |
| | P | 4.9 | 36.4 | 0.01 | 2.4 | 8.4 | 0.1 | 26.1 | 28.4 | 25.2 |
| | Q | 8.6 | >100 | 1.5 | 3.1 | 61.5 | 1.6 | 11.9 | 18.7 | 8.2 |
| | $f_V$ | 0.03 | 0.05 | 0.02 | 0.03 | 0.05 | 0.02 | 0.05 | 0.2 | 0.004 |
| | $f_I$ | 0.02 | 0.05 | 0.0 | 0.02 | 0.08 | 0.0 | 0.07 | 1.4 | 0.04 |
| 10 | V | 2.5 | 5.0 | 2.3 | 2.6 | 4.1 | 2.4 | 2.6 | 3.9 | 2.3 |
| | I | 2.5 | 5.1 | 0.9 | 1.6 | 3.4 | 0.9 | 1.7 | 3.5 | 1.0 |
| | P | 4.8 | 38.7 | 0.01 | 2.4 | 7.6 | 0.1 | 2.3 | 9.8 | 0.1 |
| | Q | 7.0 | >100 | 1.5 | 3.0 | 29.4 | 1.5 | 3.5 | 50.3 | 1.6 |
| | $f_V$ | 0.01 | 0.02 | 0.003 | 0.01 | 0.03 | 0.003 | 0.02 | 0.05 | 0.003 |
| | $f_I$ | 0.01 | 0.03 | 0.0 | 0.02 | 0.07 | 0.0 | 0.02 | 0.2 | 0.0 |
| 30 | V | 1.1 | 4.9 | 0.8 | 1.2 | 2.8 | 1.0 | 1.3 | 2.7 | 1.0 |
| | I | 2.9 | 5.5 | 0.3 | 1.8 | 3.0 | 0.4 | 1.7 | 3.2 | 0.5 |
| | P | 4.9 | 36.4 | 0.01 | 2.5 | 8.5 | 0.1 | 2.3 | 7.6 | 0.1 |
| | Q | 8.5 | >100 | 1.5 | 3.0 | 61.0 | 1.5 | 3.5 | 57.0 | 1.6 |
| | $f_V$ | 0.003 | 0.02 | 0.0 | 0.004 | 0.02 | 0.0 | 0.01 | 0.03 | 0.0 |
| | $f_I$ | 0.004 | 0.04 | 0.0 | 0.01 | 0.05 | 0.0 | 0.01 | 0.2 | 0.0 |
| 52 | V | 0.8 | 3.5 | 0.3 | 1.0 | 2.7 | 0.7 | 1.0 | 2.4 | 0.7 |
| | I | 3.1 | 5.6 | 0.5 | 1.8 | 3.4 | 0.4 | 1.9 | 3.2 | 0.5 |
| | P | 4.8 | 38.7 | 0.01 | 2.4 | 7.6 | 0.1 | 2.3 | 9.7 | 0.1 |
| | Q | 7.0 | >100 | 1.5 | 3.0 | 29.4 | 1.5 | 3.5 | 50.2 | 1.6 |
| | $f_V$ | 0.003 | 0.01 | 0.0 | 0.004 | 0.02 | 0.003 | 0.004 | 0.04 | 0.003 |
| | $f_I$ | 0.002 | 0.04 | 0.003 | 0.01 | 0.04 | 0.0 | 0.01 | 0.06 | 0.0 |

TABLE IV
STATISTICS OF $\Delta x_{t,M}$ FOR NORMAL OPERATION WITH HARMONICS, GIVEN IN % OF RMS VALUE OF CORRESPONDING VARIABLE, FOR CONSTANT (0), INCREASING (+) AND DECREASING (−) LOAD IMPEDANCE

| $f_s$ (kHz) | $\Delta L$ | 0 | | | + | | | − | | |
|---|---|---|---|---|---|---|---|---|---|---|
| | | Avg | Max | Min | Avg | Max | Min | Avg | Max | Min |
| 5 | V | 37.5 | 43.2 | 37.0 | 37.9 | 40.2 | 37.2 | 9.0 | 11.3 | 8.3 |
| | I | 8.7 | 26.0 | 5.4 | 15.3 | 50.0 | 8.4 | 44.5 | >100 | 8.0 |
| | P | 5.0 | 36.7 | 0.1 | 2.6 | 8.9 | 0.2 | 5.1 | 6.2 | 4.8 |
| | Q | 8.7 | >100 | 1.5 | 4.2 | 78.5 | 1.9 | 2.8 | 4.8 | 1.2 |
| 10 | V | 18.2 | 22.3 | 17.7 | 18.8 | 22.2 | 18.0 | 18.8 | 21.5 | 17.0 |
| | I | 6.5 | 18.0 | 2.4 | 6.0 | 26.2 | 3.1 | 4.9 | 12.2 | 2.4 |
| | P | 4.8 | 39.0 | 0.1 | 2.5 | 7.9 | 0.1 | 2.7 | 10.6 | 0.2 |
| | Q | 7.1 | >100 | 1.5 | 4.0 | 40.8 | 1.9 | 4.0 | 57.6 | 1.8 |
| 30 | V | 7.3 | 13.3 | 6.8 | 7.7 | 10.2 | 6.8 | 7.6 | 10.7 | 5.8 |
| | I | 6.2 | 11.2 | 1.0 | 3.9 | 6.4 | 1.1 | 3.9 | 6.6 | 1.0 |
| | P | 5.0 | 36.7 | 0.1 | 2.6 | 8.9 | 0.1 | 2.6 | 8.3 | 0.1 |
| | Q | 8.6 | >100 | 1.5 | 3.9 | 77.5 | 1.8 | 3.9 | 70.9 | 1.8 |
| 52 | V | 4.4 | 8.7 | 3.9 | 4.9 | 8.8 | 3.8 | 4.7 | 7.6 | 3.0 |
| | I | 6.3 | 11.0 | 0.8 | 3.0 | 6.8 | 1.0 | 4.0 | 6.5 | 0.9 |
| | P | 4.8 | 39.0 | 0.1 | 2.5 | 7.9 | 0.1 | 2.6 | 10.4 | 0.2 |
| | Q | 7.0 | >100 | 1.5 | 3.9 | 40.6 | 1.8 | 3.9 | 57.6 | 1.8 |

Tables II and IV, no statistics of metric $\Delta x_{t,M}$ are noted for frequency. The reason is that frequency is averaged over multiple electric cycles (See Section III.B), hence there is no point error (and for that matter maximum point error).

Indicatively, some statistics are here read for the ease of the reader. From Table I, for the scenario of increasing load at $f_s$=10kHz of the methodology with no harmonics in the system, the maximum of the average errors $\widetilde{\Delta x}_t$ of the active power of the digital twin over all simulations for this scenario is 7.6% with regards to the RMS of the actual active power of the T/F. From Table IV, for the scenario of constant load at $f_s$=30kHz of the method with harmonics in the system, the average of the maximum errors $\Delta x_{t,M}$ of the current of the digital twin over all simulations for this scenario is 6.2% with regards to the RMS of the actual active power of the T/F. For a load increase, the MV side waveforms of the digital twin at $f_s$=52kHz and the actual T/F are given in Fig. 4, to depict the performance of the method.

*2) System Faults, Unbalanced Loading and Tap-Changing Operations:* The scenarios of system faults are presented first. In Fig. 5 the phase voltages and line currents, at the occurrence of a LG fault on the MV side of a Dy11 T/F grounded on its LV side and a coil grounding the neutral of the voltage source are presented. The sampling rate of the digital twin methodology is at $f_s$=30kHz. The coil emulates the Peterson coil connected to the substation upstream from distribution T/F. The digital twin methodology fails to calculate *only* phase voltages of the MV side of the T/F by phase and magnitude. The method fails also *only* for phase voltages (but for different magnitudes and phase shifts) if:
- the fault is LLG or LG *and*
- the T/F is connected as Dy (regardless of grounding) or Yy with only one or neither of the two sides grounded *and*
- the substation is grounded at its MV side.

Phase voltages of the MV side of the T/F are also the *only* ones miscalculated at the occurrence of any type of fault on the LV side of the T/F. For the ease of the reader the rest of these plots are omitted.

The phase voltages and line currents, at the occurrence of a LG fault on the MV side of a Dy11 T/F grounded on its LV side and without grounding of the neutral of the voltage source are presented in Fig. 6. The sampling rate of the digital twin method is at $f_s$=30kHz. The absence of grounding of the voltage source is equivalent to the substation upstream of the distribution T/F not being grounded on its MV side. In this case, the digital twin phase voltages of the MV side of the T/F and the actual ones compare successfully. The methodology succeeds similarly, if:
- the fault is LL and regardless of any topology specifics for either the T/F or the substation *or*
- the fault is either LG or LLG, the substation is not grounded on its LV side and regardless of the topology specifics for the distribution T/F *or*
- the fault is LG or LLG and the T/F is of the Yy topology grounded on both its sides and regardless of the topology specifics of the substation.

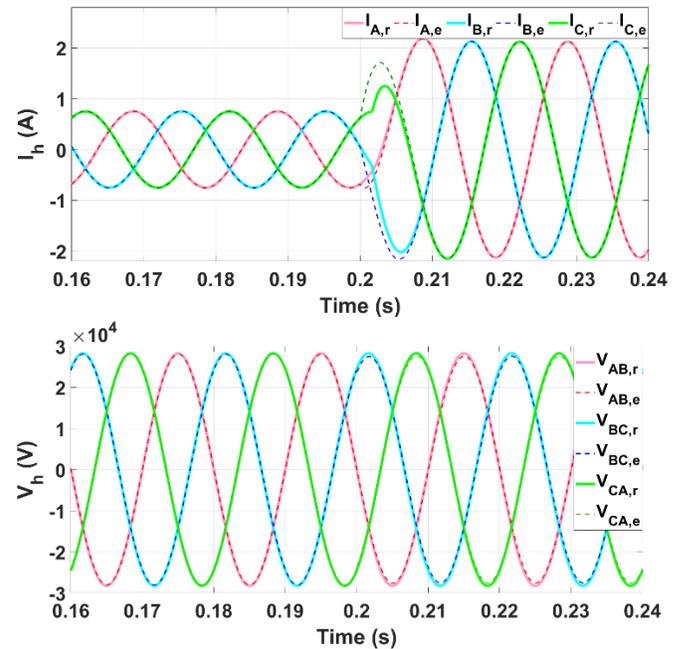

Fig. 4. MV side current and line-to-line voltage waveforms of the digital twin (subscript $e$, $f_s$=52kHz) and of the actual T/F (subscript $r$) at a load increase.

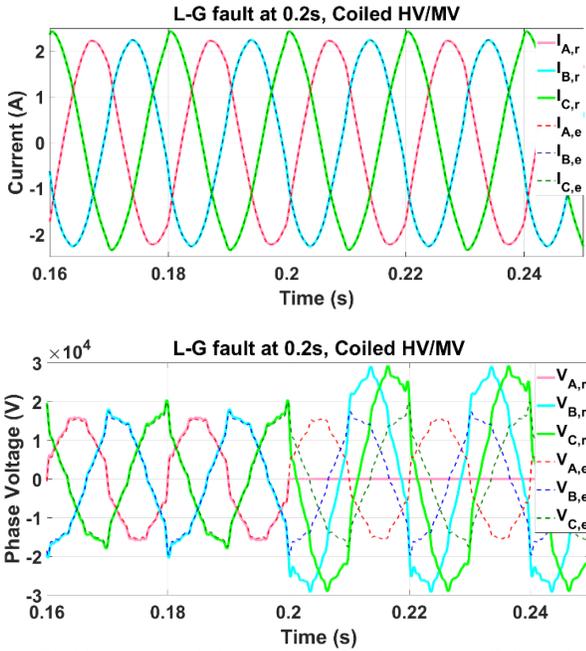

Fig. 5. MV side current and phase voltage waveforms of the digital twin (subscript $e$, $f_s$=30kHz) and of the actual T/F (subscript $r$) at a LG fault with a grounded voltage source supply.

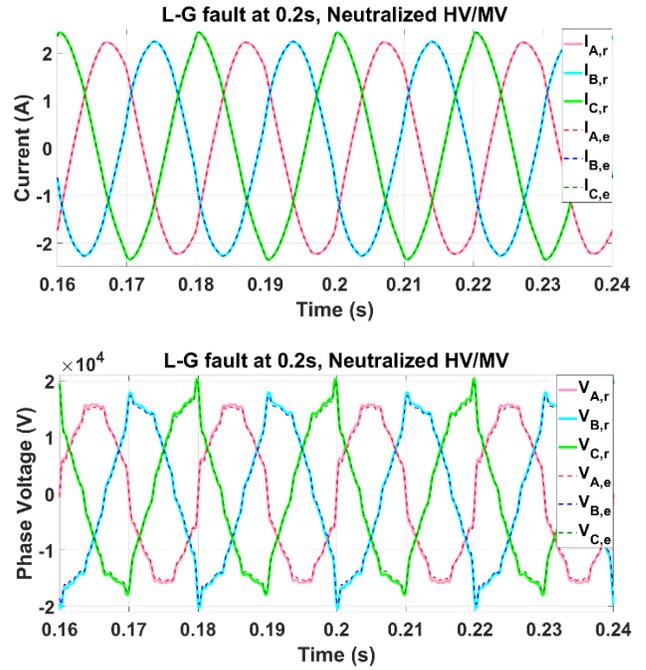

Fig. 6. MV side current and phase voltage waveforms of the digital twin (subscript $e$, $f_s$=30kHz) and of the actual T/F (subscript $r$) at a LG fault with a non-grounded voltage source supply.

For the ease of the reader the rest of these plots are omitted.

It is stressed again that the digital twin method calculates correctly all line-to-line voltages, line currents, active and reactive powers of the MV side of the T/F under any type of fault, regardless of T/F and DS grounding topologies.

Following, various cases of asymmetrical voltage magnitude of the source and asymmetrical load impedances were tested. In all cases, the digital twin methodology calculated correctly all voltage and current waveforms of the T/F MV side. As for tap-changing actions, the accuracy of the digital twin remains unaffected. This was expected as any tap-changing action only affects the parameters of the digital twin model, but not its mathematical behavior, and, thus, its accuracy. For the ease of the reader all plots of these tests are omitted.

*3) Filtering Effect of T/F and Digital Twin on Harmonics:* The Bode diagrams of the circuits representing the digital twin and the actual T/F are first plotted. The respective transfer functions of circuit (b) (digital twin) and circuit (a) (actual T/F) in Fig. 1 are, first, calculated. Load equal to 10% and 100% of the rated apparent power of the T/F at 0.8 lag power factor has been tested, to assess how the T/F circuit models (acting as filters) suppress harmonic components (even beyond those mentioned in the standard [34]). Both loading tests had similar results and the case of 100% loading is shown in Fig. 7.

In Fig. 8 the Fourier transform of voltage and current waveforms of the MV side of the digital twin ($f_s$= 30kHz) and the actual T/F are given. A load increase from 10% to 100% loading has been tested. Given how the suppression of the higher harmonics between the actual and the digital twin of the MV sides of the T/F is substantially different beyond the 20$^{th}$ harmonic (see Fig. 7), the detail in the Fourier transform of voltage is focused on the interval of 20$^{th}$-25$^{th}$ harmonics.

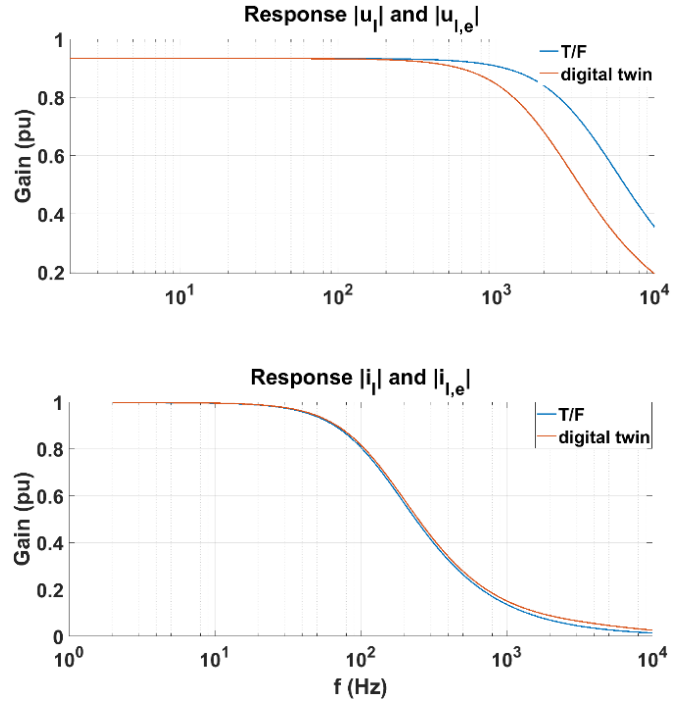

Fig. 7. Bode diagrams of load voltage and current magnitude response to electric frequency components up to 10kHz for nominal loading of the T/F for the actual and the digital twin circuit models.

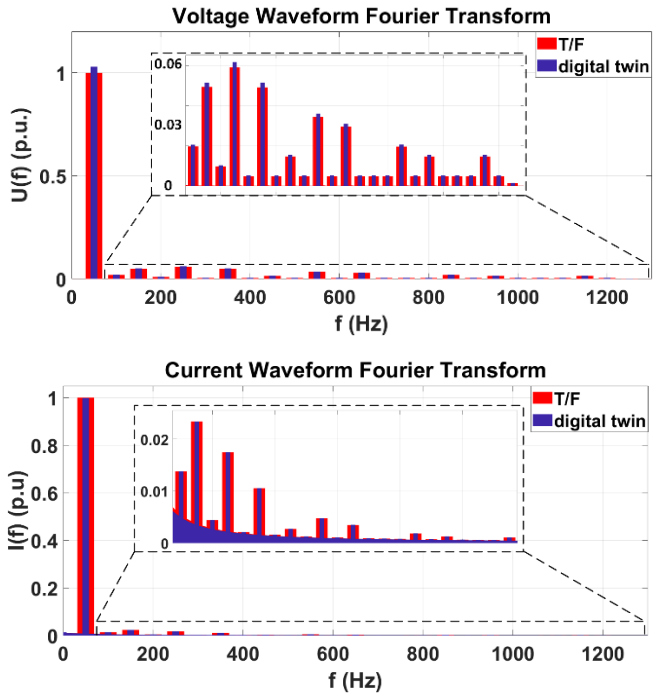

Fig. 8. Fourier transform of voltage and current waveforms of MV side for actual and digital twin ($f_s$=30kHz) of a T/F under a load increase.

### B. Discussion

*1) Statistical Testing under Normal Operation:* To assess the performance of the methodology under normal operation (Tables I-IV) both at the presence and absence of harmonics, Fig. 9 is here presented. The figure shows the envelope of maximum and minimum of the average error metric $\widetilde{\Delta x_t}$ around the average of the said metric of the voltage waveform of the digital twin from that of the actual T/F MV sides for decreasing load (as taken from Table I). It is clear that the average is considerably closer to its minimum value than its maximum. This can be noticed for all values, for both metrics ($\widetilde{\Delta x_t}$ and $\Delta x_{t,M}$) in Tables I-IV. This observation translates as that *the average behavior of the here proposed digital twin methodology is more likely to be represented by its best performance rather than calculate waveforms that deviate considerably from the ones of the actual T/F*. This observation is even more substantial for the statistics of the maximum error metric in $\Delta x_{t,M}$. The *averages of metric $\Delta x_{t,M}$ can be used to define an interval around the averages of error metric $\widetilde{\Delta x_t}$ instead of the standard deviation of the latter, representing, thus, a more realistic error interval for the waveform calculations of the digital twin methodology.* Fig. 10 shows exactly that, as the expected error interval for all waveforms of the digital twin of the MV side of the T/F for the case of constant load. For either increasing or decreasing load the respective graphs are similar (see Tables I-IV for the statistics) and have been omitted.

In terms of absolute values, let it be first reminded that instrument voltage transformers are classified according to their accuracy errors which span the interval 0.1-3% over the voltage RMS value [17]. It may be seen that at the absence of harmonics and for $f_s \geq$ 30kHz even the average maximum errors fall in this interval for the proposed digital twin methodology. At the presence of harmonics, the digital twin methodology can have accuracy performance comparable to that of the voltage T/F only by the average metric at $f_s \geq$ 10kHz. Fig. 10 confirms this observation as error tends to decrease as $f_s$ increases. With limited calculation errors of all MV waveforms as $f_s$ increases, the complexity introduced by expressing T/F resistances as functions of temperature (see Section II.A) can be avoided.

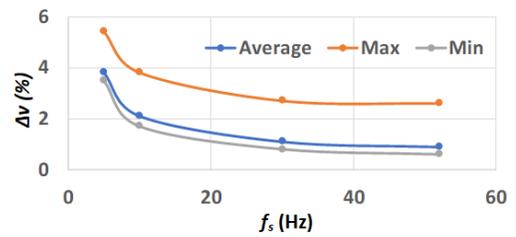

Fig. 9. Statistics of average error metric as of $\widetilde{\Delta x_t}$ for voltage of the MV digital twin of the T/F at the absence of harmonics.

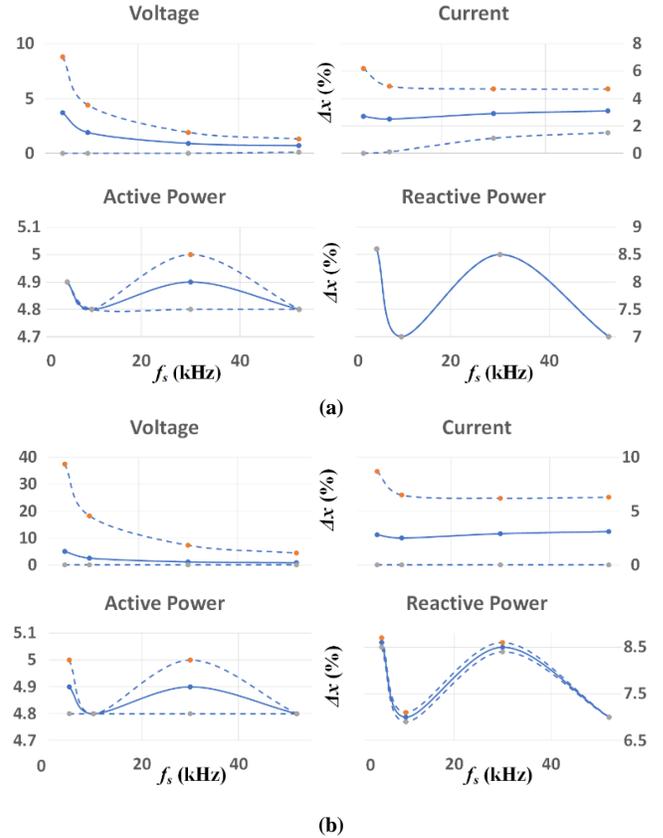

Fig. 10. Expected error interval of waveform calculations by the digital twin of the MV side of T/F compared to the actual ones at the absence (a) and presence (b) of harmonics and with constant load connected to the T/F.

Some particularly high maximum errors for both metrics for the active and reactive powers may be noted in Tables I-IV. Those errors are noted to occur only at loading of the T/F less than 5% of its nominal, thus, are not of concerning nature. Also, it is reminded that $\widetilde{\Delta x_t}$ and $\Delta x_{t,M}$ are normalized over the RMS of the respective waveform. Hence, if the normalization of $\widetilde{\Delta x_t}$ and $\Delta x_{t,M}$ was over the nameplate power instead of the RMS of the waveform, they would be negligibly small, i.e. less than 4%.

As for the frequency calculated by the digital twin, the averages of the average error metric $\widetilde{\Delta x_t}$ is about the same regardless on whether the frequency is calculated by the voltage or the current waveform (Tables I and III), with a slight favorite of the voltage. Nevertheless, from the maximum errors of the metric it is made clear that the voltage waveform should be preferred to estimate system frequency with better accuracy.

TABLE V
METRICS OF $\widetilde{\Delta x}_t$ AND $\Delta x_{t,M}$ BETWEEN THE DIGITAL TWIN (AT 13.2 kHz SAMPLING RATE) AND THE ACTUAL WAVEFORMS OF THE MV-SIDE OF A DISTRIBUTION TRANSFORMER IN SWITZERLAND, GIVEN IN % OF RMS VALUE OF CORRESPONDING VARIABLE

|       | Dataset 1 | | Dataset 2 | | Dataset 3 | | Dataset 4 | |
|-------|-----------|-----------|-----------|-----------|-----------|-----------|-----------|-----------|
|       | $\widetilde{\Delta x}_t$ | $\Delta x_{t,M}$ | $\widetilde{\Delta x}_t$ | $\Delta x_{t,M}$ | $\widetilde{\Delta x}_t$ | $\Delta x_{t,M}$ | $\widetilde{\Delta x}_t$ | $\Delta x_{t,M}$ |
| $V_R$ | 3.8 | 7.1 | 2.9 | 5.4 | 2.9 | 5.0 | 2.9 | 6.1 |
| $V_S$ | 4.7 | 8.7 | 3.6 | 7.0 | 3.7 | 6.1 | 3.9 | 8.2 |
| $V_T$ | 3.9 | 7.3 | 2.9 | 5.5 | 3.0 | 5.5 | 3.1 | 6.4 |
| $I_R$ | 12.7 | 38.5 | 14.6 | 27.9 | 12.5 | 25.4 | 15.7 | 55.4 |
| $I_S$ | 15.8 | 38.0 | 16.3 | 29.8 | 13.7 | 26.2 | 18.9 | 51.7 |
| $I_T$ | 19.7 | 61.6 | 19.2 | 34.7 | 16.8 | 32.2 | 26.6 | 80.8 |
| $P$   | 20.2 | 24.6 | 20.3 | 21.9 | 18.2 | 19.9 | 24.7 | 28.0 |
| $Q$   | 5.8  | 21.3 | 6.6  | 13.3 | 4.3  | 11.4 | 5.5  | 12.3 |

Harmonics affect the accuracy of the voltage waveform outputted by the digital twin much more than any other value (current, powers, frequency). Comparing Tables I and III, for the same type of load change and $f_s$, one can notice that statistics for all values but for voltage are similar. Especially for voltage, the metrics of the average errors are greater in Table III that tested for voltage supply with harmonics. This remark can be explained by the analysis of transfer function responses with the Bode diagrams in Fig. 7, as also the circuit itself. The shunt elements of the T/F pose practically infinite impedance to the currents through it. Hence, the voltage drops caused by the series elements (impedances with $S$, $1$, $2$ subscripts in Fig. 1) are those that affect the accuracy of the methodology, while there is no current divider to affect the accuracy from the perspective of currents. In Fig. 7 it may be noted that higher frequency components of voltage are damped more substantially by the digital twin model compared to the actual of the T/F. This, in turn, means that the digital twin methodology model will amplify any measurement errors (due to sampling in the context of this study) more than what would happen if the actual T/F model was used as the digital twin. However, as from the Fourier transform analysis in Fig. 8, it is clear that the energy contents of voltage of the actual and digital twin models of the MV side of the T/F for the frequencies that are most affected (as shown by the Bode diagrams in Fig. 7) are almost identical for a $f_s$ of 30kHz. This also confirms findings noted in the previous paragraph. In other words, preferring the actual T/F model (circuit (a) in Fig. 1) instead of the one used (circuit (b) in Fig. 1) will not improve substantially the accuracy of the digital twin, if the $f_s$ is adequately high. To conclude these observations, the accuracy of the digital twin output is unaffected by the presence of harmonics (i.e. calculates MV waveforms properly), provided that the sampling rate of the input waveforms is high enough.

*2) System Faults, Unbalanced Loading and Tap-Changing Operations:* With regards to the performance of the digital twin method for the case of system faults, the following are noted. The method relies on measurements of the LV side of the T/F, hence, unless the LV side is implicated in a phenomenon, the method will fail to calculate phase voltages. In this sense, it is here reminded that no currents flow or voltages are induced to the LV side of the *zero sequence circuit component* of Dy (regardless of grounding), YGy and Yyg T/F at the occurrence of faults on the MV side of such T/F [29]. Hence, if LG or LLG faults occur and the MV side of the DS is grounded (with Peterson coil or other), the zero sequence circuit at the MV side of the said T/F is subject to voltage and current that are not propagated to the LV side, hence the method cannot calculate phase voltages correctly.

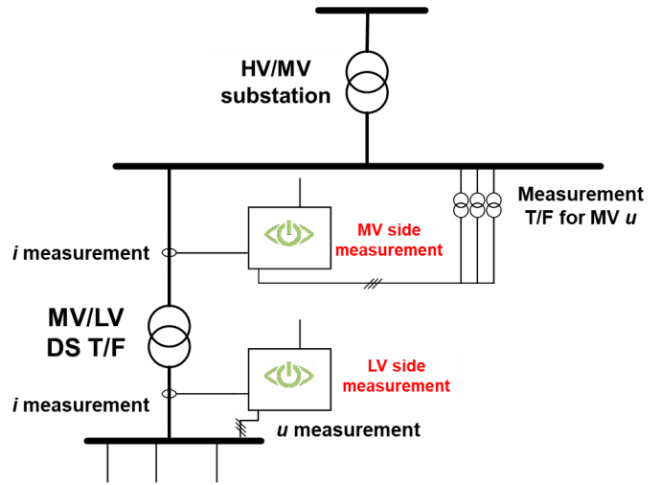

Fig. 11. Topology of the collection of field data from the actual T/F in Switzerland to assess the validity of the digital twin methodology.

The digital twin performs similarly for faults on the LV side of the T/F, too, because the voltage measurements of the method taken on the LV side of the T/F see towards the fault. I.e. due to the, typically, much smaller fault impedance, those measurements see close to zero voltage on the affected phases, hence, via (2) fail on MV side calculations. Nevertheless, this concern is not critical, since the priority of system operators in such occasions would be the LV fault (captured by the method).

## V. FIELD TEST VALIDATION OF DIGITAL TWIN OF DISTRIBUTION POWER TRANSFORMER METHODOLOGY

To assess the validity of the proposed method, current and voltage waveform measurements from an actual DS MV-LV T/F in Switzerland, are used (for T/F characteristics, see Appendix). The waveform measurements were logged with GridEye devices [33] on both its LV and MV sides in sync (see Fig. 11). GridEye, although not designed to operate as a PMU, it largely abides by the measurement requirements (see Appendix) of the respective standard [36].

LV waveform measurements of the actual T/F are inputted to the digital twin method, and the output of calculated MV waveforms of the digital twin is compared to the respective MV waveform measurements of the said T/F. Four sets of waveform data, of about 10 minutes each, sampled at 13.2kHz are used. To assess the performance of the method Table V presents the metrics defined by $\widetilde{\Delta x}_t$ and $\Delta x_{t,M}$ as the average and maximum errors of the waveform outputted by the digital twin method normalized over the RMS of the actual waveform of the T/F.

As the sampling rate of the input LV waveforms is 13.2kHz and there is ample harmonic content as the data sets show (see Fig. 12), the results in Table V will be compared to those of Tables III and IV for sampling rates of 10kHz and 30kHz.

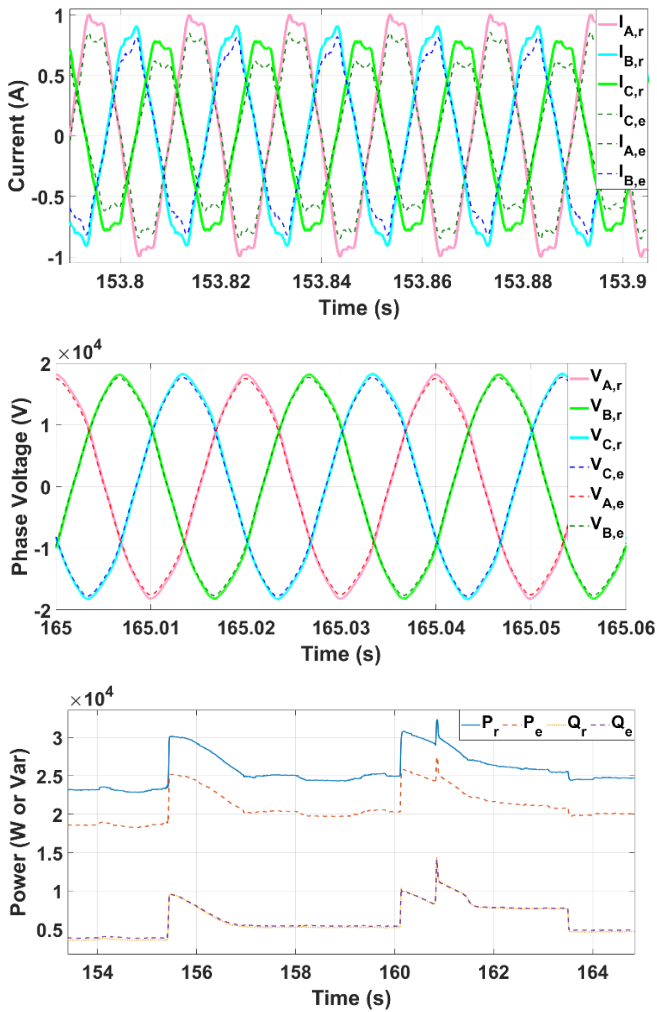

Fig. 12. MV side current and phase voltage waveforms, and active and reactive power of the digital twin (subscript $e$, $f_s$=13.2 kHz) and of the MV measurement data of the actual T/F in Switzerland (subscript $r$).

As it may be noted all voltage, active and reactive power metrics are within the intervals of the statistical testing with the synthetic/simulation data. The only exception is with the metrics of the currents, which are greater than the maxima of both $\widetilde{\Delta x_t}$ and $\Delta x_{t,M}$. Since the metrics of voltages and powers are within the intervals of the statistical results collected from the simulations with synthetic data and the current is a function of them, the noted error metrics of current are a result of how these metrics are normalized over the RMS of the respective waveform. In fact, the T/F loading across all four datasets was particularly low (less than 5% of the nameplate 630kVA); i.e. if the normalization of the error metrics was over the nominal current instead the RMS of the waveform, the errors would be negligibly small, i.e. no more than 4%. Similar observations were made for the high error metrics of the active and reactive power of the digital twin for the statistical testing with synthetic data (see Section IV.B).

The generally positive behavior of the digital twin method, as shown by the above error metrics analysis, is corroborated by the comparison of the waveforms of the MV-side of the T/F between the output of the digital twin and the datasets of the actual T/F in Fig. 12. The field data clearly shows that the digital twin methodology can accurately calculate the MV-side behavior of a T/F based on LV-side waveform input.

## VI. CONCLUSION

This work proposes a method to monitor the MV-side waveforms of DS T/F, to enhance the visibility of operators into DS. The method is the digital twin of the MV side of a DS T/F. It has high calculation accuracy and captures most system faults and harmonics content. The accuracy of the digital twin increases as the sampling rate of the input LV waveforms increases, and is comparable to that of an instrument T/F.

Next step in the assessment of the proposed technique is to deploy the method on a DS monitoring device [33] to a MV-LV T/F and assess its field performance. Thus, the technique will also be realistically assessed in terms of installation costs and disruption. Practically, the digital twin method is expected to offer much visibility of the MV DS to DS operators. Moreover, the health of the T/F can be monitored through its performance.

A first research extension that can be explored on the T/F digital twin method, is its improvement in cases of compensated MV DS ground faults, if not both the T/F primary and secondary windings are grounded. The proposed digital twin T/F set-up can calculate accurately MV current flows and line-to-line voltages under all MV system fault conditions, but fails to calculate the phase voltages (indicator of nature of fault) in the aforementioned specific cases. Another extension can be the application of the method to MV to high voltage T/F to monitor the transmission system with reduce costs and grid disruptions.

## VII. APPENDIX

### A. Simulation Testing on MATLAB

MV-LV DS 3-phase T/F parameters:
$S$ = 50 kVA, $V_1/V_2$ = 400V/20kV, $R_1$ = 0.0075 pu, $L_1$ = 0.02 pu, $R_2$ = 0.0075 pu, $L_2$ = 0.02 pu, $R_m$ = 500 pu, $L_m$ = 500 pu.

LV 3-phase lumped varying load parameters:
$R_L$ = [0.75-5.25] Ω, $L_L$ = [1.5-17] mH.

MV 3-phase line parameters:
$R_{LN}$ = 2 Ω, $L_{LN}$ = 1 mH.

### B. Digital Twin Tests with Field Data from T/F in Switzerland

LV-MV DS 3-phase T/F parameters:
$S$ = 630 kVA, $V_1/V_2$ = 400V/20.5kV, $R_1$ = 0.0035 pu, $L_1$ = 0.0233 pu, $R_2$ = 0.0035 pu, $L_2$ = 0.0233 pu, $R_m$ = 500 pu, $L_m$ = 500 pu.

GridEye measurement accuracy characteristics:
Voltage magnitude = 0.1%,
Total Vector Error (see [36]) < 1%,
Current Magnitude = 1%,
Frequency = ±1 mHz.